\documentclass[preprint,12pt,nopreprintline]{elsarticle}

\usepackage{lmodern}
\usepackage[hidelinks]{hyperref}
\usepackage{amsmath,amssymb}
\usepackage{mathtools}
\usepackage{bm}
\usepackage{longtable}
\usepackage{pdflscape}
\usepackage{array}
\usepackage{booktabs}
\usepackage{enumitem}
\usepackage{graphicx}
\usepackage{fancyvrb}
\usepackage{xurl}
\usepackage{xcolor}
\usepackage{tikz}
\usepackage[section]{placeins}
\usetikzlibrary{arrows.meta,positioning}
\newcommand{\PackageVersion}{0.3.3}
\newcommand{\SoftwareReleaseURL}{https://github.com/lennonshikhman/diagnostics_for_physics/releases/tag/v0.3.3}

\hypersetup{pdftitle={PDECheck: A Diagnostic Software Suite for Auditing Learned PDE Simulators},pdfauthor={Lennon J. Shikhman}}

\begin{document}

\begin{frontmatter}

\title{PDECheck: A Diagnostic Software Suite\\for Auditing Learned PDE Simulators}

\author[gt,fit]{Lennon J. Shikhman\corref{cor1}}
\ead{lj@shikhman.net}
\cortext[cor1]{Corresponding author}

\affiliation[gt]{organization={College of Computing, Georgia Institute of Technology},
            city={Atlanta},
            state={GA},
            country={USA}}

\affiliation[fit]{organization={College of Engineering and Science, Florida Institute of Technology},
            city={Melbourne},
            state={FL},
            country={USA}}

\begin{abstract}
PDECheck audits learned PDE simulators through callable propagators or saved predictions. It evaluates relative \(L^2\) error, temporal composition, finite-difference generators, and equation-specific physical residuals, retaining horizons and missing prerequisites. Analytic and manufactured defects verify the kernels. A heat-equation sweep quantifies generator sensitivity to timestep, frequency, grid, and precision. A three-equation, five-seed FNO study compares supervised, PINO-style, and composition-regularized training. 
An accuracy-matched advection comparison has state errors below 0.3\% and a semigroup-error ratio of 1.48 (95\% interval 1.34--1.62). The composition penalty lowers semigroup error but increases composed-state and longer-rollout prediction error. \ignorespaces

 A time-reparameterization check produces identical low-error endpoints with different compositions. Applications to five external equation families exercise the readers and physical contracts on coarse grids. The release includes reproducibility artifacts and measured CPU/CUDA costs.
\end{abstract}


\end{frontmatter}

\section{Motivation and significance}
\label{sec:introduction}

A time-conditioned PDE surrogate can return different states for one call of duration \(t+s\) and two calls of durations \(s\) and \(t\). Endpoint prediction error does not measure this composition defect, mass conservation, or density positivity.

The underlying tests are established: finite-volume balance \cite{leveque2002finite}, conservation \cite{hairer2006geometric}, and the autonomous semigroup identity \cite{pazy1983semigroups,engel2000oneparameter}. PDECheck makes them callable audits of trained models, using declared state variables and boundary conditions. Saved predictions support endpoint tests; callable adapters also restart the model from intermediate or modified states.

Analytic defects and controlled FNO comparisons separate composition consistency from prediction accuracy. External applications use PDEBench \cite{takamoto2022pdebench}, APEBench \cite{koehler2024apebench}, and The Well \cite{ohana2024thewell} across five equation families.

\section{Software metadata and availability}
\label{sec:software_metadata}

Table~\ref{tab:software-metadata} identifies the release; \ref{app:reproduction} describes its reproducibility artifacts.

\begin{table}[htbp]
\centering
\caption{Software metadata for PDECheck}
\label{tab:software-metadata}
\small
\setlength{\tabcolsep}{6pt}
\begin{tabular}{>{\raggedright\arraybackslash}p{0.32\linewidth}>{\raggedright\arraybackslash}p{0.58\linewidth}}
\toprule
Metadata field & Value \\
\midrule
Software name & PDECheck / \texttt{pdecheck} \\
Repository & \url{https://github.com/lennonshikhman/diagnostics_for_physics} \\
Primary use & Post hoc auditing of learned PDE simulators and neural surrogate time propagators \\
Version / language & \PackageVersion\ / Python 3.10+ \\
Versioned release & \href{\SoftwareReleaseURL}{v\PackageVersion} \\
Input data & Reference trajectories, learned-model predictions or callable propagators, equation metadata, and diagnostic configuration \\
Main outputs & Per-diagnostic scalar summaries, equation-family diagnostic panels, and tables comparing model variants \\
Supported model classes & Architecture-agnostic; applicable to neural operators, convolutional surrogates, autoregressive emulators, and saved prediction arrays \\
License & MIT License  \\
\bottomrule
\end{tabular}
\end{table}

\section{Software description}
\nopagebreak[4]

The panel combines reference trajectories \(\{u^m_0,u^m_1,\ldots,u^m_T\}_{m=1}^M\), a propagator \(\widehat{\Phi}_{\theta,t}:X_h\to X_h\), and equation metadata specifying physical channels, time spacing, grid, and boundary conditions. Adapters expose time-conditioned models, fixed-step predictors, or saved arrays. Diagnostics use decoded physical variables. Reports store per-sample values and aggregation axes; Section~\ref{sec:api} gives the API and failure rules.

\subsection{Architecture and audit algorithm}
\label{sec:architecture}
Figure~\ref{fig:architecture} and the pseudocode show the audit sequence.

\begin{figure}[htbp]
\centering
\begin{tikzpicture}[
 node distance=7mm,
 box/.style={draw, rounded corners=2pt, align=center, text width=0.83\linewidth, inner sep=6pt, font=\small},
 arrow/.style={-{Latex[length=2mm]}, thick}]
\node[box] (input) {Reference states and equation metadata\\Callable, fixed-step adapter, or saved predictions};
\node[box, below=of input] (validate) {Validate shapes, physical channels, precision, grid, and time\\Training also verifies split identities and provenance};
\node[box, below=of validate] (availability) {Determine applicability from the state and physical contract\\Missing prerequisites: unavailable value and explicit reason};
\node[box, below=of availability] (predict) {For each horizon: direct prediction\\Actual composition, paired perturbation, and scaling calls when supported};
\node[box, below=of predict] (kernels) {Evaluate per-sample diagnostic kernels\\State, composition, generator, functional, balance, and constraints};
\node[box, below=of kernels] (report) {Reject nonfinite applicable results; retain counts and horizon pairs\\Return reports, CSV/JSON, and source/data provenance};
\draw[arrow] (input) -- (validate);
\draw[arrow] (validate) -- (availability);
\draw[arrow] (availability) -- (predict);
\draw[arrow] (predict) -- (kernels);
\draw[arrow] (kernels) -- (report);
\end{tikzpicture}
\caption{PDECheck architecture, version \PackageVersion. The TikZ source was developed during revision with OpenAI Codex (OpenAI; GPT-6.1 Sol, GPT-6 Sol, GPT-6 Astra).}
\label{fig:architecture}
\end{figure}

The audit uses the first positive requested horizon as the generator anchor. Fixed-step adapters supply integer-step splits for composition.
\noindent\begin{minipage}{\linewidth}
\begin{Verbatim}[fontsize=\footnotesize,frame=single]
AUDIT(initial, references, propagator, equation_metadata):
    validate finite physical states, shapes, channels, and grid
    horizons <- sorted positive reference horizons
    for t in horizons:
        direct[t] <- checked propagator(initial, t)
        report[t] <- applicable pair diagnostics(initial,
                         references[t], direct[t], elapsed_time=t)
        if an actual composition can be evaluated:
            composed <- restart at an intermediate state
            report[t].semigroup <- norm(composed - direct[t])
                                      / (norm(direct[t]) + epsilon)
            annotate fixed-step equality by construction
        otherwise: record unavailable and its reason
        if t exceeds the first horizon:
            compare generators at (first_horizon, t)
        if paired perturbed references are supplied:
            propagate the perturbed state and compare responses
        if a valid scaling contract is supplied:
            propagate the transformed state and compare
        reject failed applicable values; retain all counts
    return per-horizon reports and labelled pooled summary
\end{Verbatim}
\end{minipage}

\section{Background and related work}
\label{sec:related_work}

Neural operators learn maps between function spaces \cite{kovachki2023neuraloperator}: DeepONet uses branch and trunk networks \cite{lu2021deeponet}, and FNO uses spectral convolutions \cite{li2021fno}. Physics-informed learning adds equation residuals \cite{raissi2019pinn,karniadakis2021physics}, as in PINO \cite{li2024pino} and physics-informed DeepONet \cite{goswami2023pidon}. Group-equivariant Fourier layers encode symmetries \cite{helwig2023group}; Cellular Sheaf Neural Operators encode compatibility through cell-complex incidence, coboundary, and flux structure \cite{shikhman2026cellularsheafneuraloperators}.

Semigroup tests compare direct and composed evolution \cite{shikhman2026semigroup}; Jacobian spectral diagnostics examine local response \cite{gao2026spectralaudit}. Spectral bias \cite{rahaman2019spectral} and FNO discretization error \cite{lanthaler2025discretizationerrorfourierneural} introduce frequency and grid dependence. Documented PINN \cite{krishnaprayan2021pinnfailures} and neural-operator failures \cite{shikhman2026diagnosing} motivate measurements beyond fitting loss. Boundary-indexed operator families distinguish operators with different boundary conditions \cite{shikhman2026one}, which PDECheck records in its metadata.

\subsection{Relationship to existing benchmark frameworks}
\label{sec:framework}
Table~\ref{tab:frameworks} compares the existing frameworks \cite{takamoto2022pdebench,koehler2024apebench,ohana2024thewell,wellmetrics} with PDECheck's callable temporal tests and named physical residuals. Users can audit existing models without adopting its training runner.

\begin{table}[htbp]
\centering
\caption{Data, training, and evaluation roles of the suite and the benchmark frameworks, from the cited publications and documentation.}
\label{tab:frameworks}
\small
\begin{tabular}{>{\raggedright\arraybackslash}p{0.17\linewidth}>{\raggedright\arraybackslash}p{0.35\linewidth}>{\raggedright\arraybackslash}p{0.38\linewidth}}
\toprule
Software & Established role & Relationship to this suite\\
\midrule
PDEBench & PDE data and training; state, conservation, boundary, and spectral errors & Reference data for a callable temporal and physical audit\\
APEBench & Procedural solvers, differentiable training, autoregressive rollouts & Direct-versus-composed audits for time-conditioned emulators; fixed-step equality identified as a construction property\\
The Well & Simulations, standardized fields/boundaries, spatial/spectral metrics & Field metadata configure residuals and identify missing prerequisites\\
PDECheck & Post hoc tests of callables, fixed-step adapters, and saved arrays & Temporal and physical residuals with applicability, failure handling, and provenance\\
\bottomrule
\end{tabular}
\end{table}

\section{Diagnostic methods implemented by the suite}

Each kernel returns a value per sample, with Euclidean norms over flattened channel and spatial axes. Denominators use \(\varepsilon=10^{-8}\); energy densities use spatial means on the uniform grid. Section~\ref{sec:api} specifies aggregation and numerical validity.

\subsection{Relative \texorpdfstring{\(L^2\)}{L2} Error}

For propagation from \(u^m_j\) to \(u^m_{j+k}\), relative state error is
\[
    E_{L^2}
    =
    \frac{
        \left\|
        \widehat{\Phi}_{\theta,k\Delta t}(u^m_j)
        -
        u^m_{j+k}
        \right\|_2
    }{
        \left\|u^m_{j+k}\right\|_2+\varepsilon
    }.
\]

\subsection{Semigroup Error}
\label{sec:semigroup}

For an autonomous well-posed evolution equation, the solution operators form a
semigroup:
\[
    \Phi_{t+s} = \Phi_t \circ \Phi_s, \qquad \Phi_0 = I.
\]
For horizons \(a\Delta t\) and \(b\Delta t\), the composition defect is
\[
    E_{\mathrm{sg}}
    =
    \frac{
        \left\|
        \widehat{\Phi}_{\theta,b\Delta t}
        \left(
        \widehat{\Phi}_{\theta,a\Delta t}(u^m_j)
        \right)
        -
        \widehat{\Phi}_{\theta,(a+b)\Delta t}(u^m_j)
        \right\|_2
    }{
        \left\|
        \widehat{\Phi}_{\theta,(a+b)\Delta t}(u^m_j)
        \right\|_2+\varepsilon
    }.
\]
Evaluation restarts the callable from its predicted intermediate state \cite{shikhman2026semigroup}; saved arrays require an actual composed endpoint. For a fixed-step adapter \(\widehat\Phi_{n\Delta t}=F^n\), integer-step composition is an algebraic identity even for inaccurate \(F\). Reports label this construction property and mark composition at a single stored step unavailable. A physical semigroup interpretation requires a deterministic autonomous map on the complete evolution state; hidden forcing or parameters require an augmented state or nonautonomous interface.

\subsection{Generator Discrepancy}
\label{sec:generator}

The infinitesimal generator of a strongly continuous semigroup is defined by
\[
    Au = \lim_{t\downarrow 0} \frac{\Phi_t u - u}{t},
\]
on its domain \cite{pazy1983semigroups,engel2000oneparameter}. The learned finite-difference estimate at \(k\Delta t\) is
\[
    \widehat{A}_{k\Delta t}(u)
    =
    \frac{
        \widehat{\Phi}_{\theta,k\Delta t}(u)-u
    }{
        k\Delta t
    }.
\]
For two short horizons \(a\Delta t\) and \(b\Delta t\), the generator
discrepancy is
\[
    E_{\mathrm{gen}}
    =
    \frac{
        \left\|
        \widehat{A}_{a\Delta t}(u^m_j)
        -
        \widehat{A}_{b\Delta t}(u^m_j)
        \right\|_2
    }{
        \left\|
        \widehat{A}_{a\Delta t}(u^m_j)
        \right\|_2
        +
        \left\|
        \widehat{A}_{b\Delta t}(u^m_j)
        \right\|_2
        +\varepsilon
    }.
\]
This score compares local evolution estimates; even the exact map generally gives a nonzero finite-horizon value. We also report
\[
 R_{\rm gen}=\frac{\overline E_{\rm gen}^{\rm model}}
 {\overline E_{\rm gen}^{\rm ref}+\varepsilon},
\]
where bars denote sample means at the same horizon pair. The ratio is unavailable for nonfinite reference means or means no larger than \(\varepsilon\). The API pairs the first positive horizon with each later one; the benchmark runner anchors at one stored step. Section~\ref{sec:ratio_interpretation} gives interpretation and sensitivity.

\subsection{Energy-law violation and reference-functional error}
\label{sec:energy}

Many PDEs possess conserved or dissipated energy-like quantities. For a
conservative system,
\[
    E(\Phi_t u)=E(u),
\]
whereas for a dissipative system,
\[
    E(\Phi_t u)\leq E(u).
\]
For conservative systems, we measure relative energy drift,
\[
    E_{\mathrm{energy}}
    =
    \frac{
        \left|
        E\left(\widehat{\Phi}_{\theta,k\Delta t}(u^m_j)\right)
        -
        E(u^m_j)
        \right|
    }{
        \left|E(u^m_j)\right|+\varepsilon
    }.
\]
For dissipative systems, we instead measure violation of monotone decay by
\[
    E_{\mathrm{energy}}
    =
    \frac{
        \max\left\{
        0,\,
        E\left(\widehat{\Phi}_{\theta,k\Delta t}(u^m_j)\right)
        -
        E(u^m_j)
        \right\}
    }{
        \left|E(u^m_j)\right|+\varepsilon
    }.
\]
The column \texttt{energy\_violation} records conservation drift or forbidden energy increase \cite{hairer2006geometric}. Reference disagreement is separate:
\[
 E_{\rm ref}=\frac{|E(\widehat\Phi_t(u))-E(\Phi_t(u))|}
 {|E(\Phi_t(u))|+\varepsilon}.
\]
Excess dissipation can give zero law violation and nonzero \(E_{\rm ref}\). Reference-only functionals have \(E_{\rm ref}\) without an asserted energy law (Section~\ref{sec:external}).

\subsection{Integral Balance Error}
\label{sec:balance}

For \(\partial_t q(u)+\nabla\cdot F(u)=S(u)\) on control volume \(K\) \cite{leveque2002finite}, let \(M_0,M_1\) be volume-weighted sums of initial and predicted \(q\), and \(\bar B,\bar S\) the endpoint-averaged outward surface flux and volume source. For cell-centered values or cell averages and elapsed time \(\tau>0\),
\[
 E_{\rm bal}(K)=
 \frac{|M_1-M_0+\tau\bar B-\tau\bar S|}
 {|M_0|+|M_1|+\tau|\bar B|+\tau|\bar S|+\varepsilon}.
\]
Cell volumes and face areas use physical spacing. Interior face fluxes average adjacent cells; periodic faces wrap, and nonperiodic outer faces use two-cell linear extrapolation. Endpoint averaging gives trapezoidal time quadrature. One-dimensional point faces have unit weight. Periodic whole-domain flux cancels; subvolume flux remains.

Reference baselines and refinement checks quantify quadrature residuals (Section~\ref{sec:manufactured}). The API accepts subvolume slices, used for the octant audit in Section~\ref{sec:external_results}. Without fluxes, the separate \texttt{global\_conserved\_drift} measures channel-sum change (\ref{app:contracts}).

\subsection{Constraint Error}
\label{sec:constraints}

Constraints are configured and reported individually. For a scalar channel \(q\) required to be nonnegative, the implementation computes
\[
 E_{\rm pos}(q)=\frac{\|\max(-q,0)\|_2}{\|q\|_2+\varepsilon}.
\]
For a vector field \(v\) required to be solenoidal, it computes
\[
 E_{\rm div}(v)=\frac{\|\nabla_h\cdot v\|_2}
 {\|\nabla_h v\|_F+\varepsilon}.
\]
\ref{app:contracts} specifies task channels and derivative stencils, which differ from balance-face quadrature. Components remain separate; the compatibility column \texttt{constraint} is available only for a single configured constraint. Custom callbacks may supply other residuals.

\subsection{Perturbation Response Error}
\label{sec:perturbation}

Small initial-state perturbations probe the tangent map,
\[
    \Phi_t(u+\delta u)-\Phi_t(u)
    \approx
    D\Phi_t(u)\,\delta u.
\]
Given a perturbation \(\delta u\), define
\[
    R_{\mathrm{ref}}
    =
    \Phi_{k\Delta t}(u+\delta u)-\Phi_{k\Delta t}(u),
    \qquad
    R_{\mathrm{model}}
    =
    \widehat{\Phi}_{\theta,k\Delta t}(u+\delta u)
    -
    \widehat{\Phi}_{\theta,k\Delta t}(u).
\]
The perturbation response error is
\[
    E_{\mathrm{pert}}
    =
    \frac{
        \left\|
        R_{\mathrm{model}}-R_{\mathrm{ref}}
        \right\|_2
    }{
        \left\|
        R_{\mathrm{ref}}
        \right\|_2+\varepsilon
    }.
\]
This measures tangent-map disagreement for infinitesimal perturbations \cite{gao2026spectralaudit} and nonlinear response for finite amplitudes.

\subsection{Scaling-Law Error}
\label{sec:scaling}

For a declared scaling transformation \(S_\lambda\) \cite{barenblatt1996scaling}, suppose the reference satisfies
\[
    \Phi_t(S_\lambda u)
    =
    S_\lambda \Phi_{\alpha(\lambda)t}(u),
\]
The corresponding error is
\[
    E_{\mathrm{scale}}
    =
    \frac{
        \left\|
        \widehat{\Phi}_{\theta,t}(S_\lambda u)
        -
        S_\lambda
        \widehat{\Phi}_{\theta,\alpha(\lambda)t}(u)
        \right\|_2
    }{
        \left\|
        S_\lambda\widehat{\Phi}_{\theta,\alpha(\lambda)t}(u)
        \right\|_2+\varepsilon
    }.
\]
Equation metadata must supply the valid transformation and time-rescaling factor.

\section{Public interface and reproducibility}
\label{sec:api}

\subsection{Using an existing propagator}
Decoded states have shape \texttt{(batch, channels, *spatial)}; unbatched channel/spatial arrays are also accepted. The API preserves float32/float64; normalization and model-specific dtype conversion belong in the adapter. An executable example audits a callable against stored references:
\begin{Verbatim}[fontsize=\small]
from diagnostic_suite import DiagnosticConfig, DiagnosticPanel
from diagnostic_suite.propagators import CallablePropagator

panel = DiagnosticPanel(DiagnosticConfig(
    dim=1, grid_spacing=(1/64,),
    energy_type="generic_l2", energy_law="dissipative"))
result = panel.evaluate_propagator(
    initial, {0.05: ref_a, 0.2: ref_b},
    CallablePropagator(predict))
report = result.by_horizon[0.2].as_report()
\end{Verbatim}
The callable \texttt{predict(state, horizon)} must accept intermediate, perturbed, or scaled states. Saved-prediction adapters accept endpoints and separately supplied compositions.

\subsection{Numerical validity and aggregation}
\texttt{DiagnosticResult} stores per-sample values, means, population standard deviations, finite/total counts, status, applicability, and reasons, grouped by \texttt{by\_horizon} with generator pairs. \texttt{as\_report()} exposes the report. Evaluated zeros remain distinct from unavailable values, exported as JSON \texttt{null}. Invalid channels, malformed callbacks, or nonfinite applicable values raise exceptions; the CLI records failure and exits nonzero.

\subsection{Data contracts and resuming experiments}
Before normalization or fitting, the workflow checks trajectory splits: seeded 90/10 train/validation with an external test file for APEBench; seeded 80/10/10 for the selected PDEBench files; official directories for The Well. Readers verify fields and record files, trajectory indices, channels, spacing, and times. NPY arrays are memory-mapped, copying only requested states. Periodic subsampling uses a stride dividing the native grid size; spatial limits are upper bounds.

Run artifacts retain arguments, versions, source fingerprints, splits, full SHA-256 input hashes, normalizers, checkpoints, and seed/horizon metrics. Resume always verifies source, dependencies, configuration, splits, and input hashes, even without the optional non-resume hashing flag. Mismatches abort without overwriting evidence; changed configurations require fresh directories. Each invocation is recorded. Table export checks package/schema versions and rejects obsolete schemas, duplicates, failed values, or incomplete requested model/variant/seed coverage when launch arguments are supplied.

\subsection{Exact smoothing procedure}
\label{sec:smoothing}
One pass on a \(d\)-dimensional prediction is the channelwise box filter
\[
 (\mathcal Sv)_{\boldsymbol j}
 =3^{-d}\!\!\sum_{\boldsymbol r\in\{-1,0,1\}^{d}}v_{\boldsymbol j+\boldsymbol r}.
\]
The variant \(\mathcal S^2\widehat\Phi_{\theta,t}\) applies two width-three, stride-one passes per axis, without channel mixing, downsampling, or retraining. Periodic tasks wrap; nonperiodic tasks replicate edge values. Every direct or restarted call is filtered before affine decoding, which commutes with this constant-preserving stencil.

A periodic mode \(\boldsymbol\omega\) is attenuated by \(\prod_{\ell=1}^d[(1+2\cos\omega_\ell)/3]^2\), verified alongside global mass preservation. Replicated edges need not preserve sums. The CLI records \texttt{--oversmooth-passes} and uses task periodicity; wrapper construction requires an explicit boolean \texttt{periodic} argument.

\subsection{Public API details}
\label{sec:api_details}
Table~\ref{tab:api} lists the principal calls and callback shapes.

\begin{table}[htbp]
\centering
\caption{Public interfaces for decoded states. Callbacks must satisfy shape and finiteness contracts; executable examples are distributed.}
\label{tab:api}
\footnotesize
\begin{tabular}{p{0.35\linewidth}p{0.55\linewidth}}
\toprule
Interface & Contract and output\\
\midrule
\texttt{evaluate\_pair} & Initial, reference, prediction, and elapsed time; endpoint state and applicable physical diagnostics\\
\texttt{evaluate\_propagator} & Initial state, positive-horizon reference mapping, and callable; separate horizon reports and supported operator tests\\
\texttt{semigroup\_error} & Independently obtained direct and composed endpoints\\
\texttt{generator\_discrepancy} & Initial state and predictions at two distinct positive physical horizons\\
\texttt{perturbation\_response} & Four arrays: base/perturbed reference and base/perturbed prediction\\
\texttt{scaling\_error} & Scaled-input prediction and transformed base prediction for a valid declared symmetry\\
\texttt{BalanceSpec} callbacks & Quantity/source: \texttt{(B,1,*spatial)}; flux: \texttt{(B,d,*spatial)}; physical spacing and time required\\
\bottomrule
\end{tabular}
\end{table}

\section{Configured external data interfaces}
\label{sec:external}

Table~\ref{tab:applicability} gives the physical tests supported by the selected files. Manifests record dataset identifiers, generating parameters, units, and boundaries when available.

The APEBench \texttt{diff\_adv} scalar-advection scenario \cite{apebenchscenario} is named \texttt{adv2d}. Its NPY files omit physical time spacing and advection coefficients: time uses stored steps, and flux balance is unavailable. The obsolete \texttt{ns2d} name is rejected.

Height-only shallow-water data support positivity and global drift, with energy and flux unavailable. Compressible-flow energy uses primitive variables,
\[
 e=\tfrac12\rho|v|^2+\frac{p}{\gamma-1},\qquad\gamma=1.4.
\]
For MHD, the missing internal energy restricts the comparison to the reference functional \(\tfrac12\rho|v|^2+\tfrac12|B|^2\). Active matter uses the squared state norm as its reference functional.

\begin{table}[htbp]
\centering
\caption{Physical diagnostics for the configured external state variables. State and generator comparisons use reference trajectories. Composition uses callable models; paired perturbed references are absent from these data.}
\label{tab:applicability}
\footnotesize
\setlength{\tabcolsep}{3pt}
\begin{tabular}{>{\raggedright\arraybackslash}p{0.13\linewidth}>{\raggedright\arraybackslash}p{0.23\linewidth}>{\raggedright\arraybackslash}p{0.23\linewidth}>{\raggedright\arraybackslash}p{0.30\linewidth}}
\toprule
Task & Stored state & Energy interpretation & Balance / constraints \\
\midrule
ADV2D & Scalar advection & Conservative squared norm & Global scalar drift; amplitude scaling \\
SW2D & Height only & Unavailable & Global height drift; height positivity \\
Active2D & Declared scalar/vector/tensor fields & Reference squared norm & No configured balance or constraint \\
CNS3D & Density, pressure, three velocities & Total energy & Mass balance; separate density and pressure positivity \\
MHD3D & Density, velocity, magnetic field & Reference kinetic-plus-magnetic energy & Mass balance; density positivity and magnetic divergence \\
\bottomrule
\end{tabular}
\end{table}

\FloatBarrier

\section{Reproducible software validation}
\label{sec:analytic}

\subsection{Independent analytic PDE reference}
We use periodic advection--diffusion on \(x\in[0,1)\),
\[
 \partial_t u+c\partial_xu=\nu\partial_{xx}u,
 \qquad c=0.3,\quad\nu=0.02,
\]
with four smooth positive initial states indexed by
\(\phi\in\{0,0.4,0.8,1.2\}\). The closed-form reference is
\begin{align*}
 u(x,t;\phi)={}&1+0.2e^{-\nu(2\pi)^2t}
                  \sin(2\pi(x-ct)+\phi)\\
              &+0.1e^{-\nu(4\pi)^2t}
                  \cos(4\pi(x-ct)-\phi).
\end{align*}
Reference arrays use this expression; an independent FFT callable multiplies arbitrary-input Fourier coefficients by \(\exp[(-ick-\nu k^2)t]\). The audit uses 64 points, CPU float64, and horizons \(0.05,0.1,0.2\), without fitting.

Contracts specify dissipative \(E(u)=\tfrac12\langle u^2\rangle\), positivity, and mass flux \(cu-\nu u_x\) with a spectral derivative. The perturbation \(10^{-3}\cos(6\pi x)\) has an independently evaluated translated/damped reference. Scaling uses \(S_\lambda u=\lambda u\), \(\lambda=1.25\), without time rescaling.

\subsection{Controlled propagator defects}
The same initial states and PDE test the FFT map and four defects: frozen \(\widehat\Phi_t(u)=u\), clock \(\widehat\Phi_t=\Phi_{t+0.75t^2}\), excess diffusion \(4\nu\), and source leak \(\widehat\Phi_t(u)=\Phi_t(u)+0.4t\).

\begin{table}[htbp]
\centering
\caption{Generated validation output at \(t=0.2\), averaged over four initial states. The generator pair is \((0.05,0.2)\). Error columns are normalized as specified in the methods; \(R_{\rm gen}\) is interpreted relative to one. Values near machine precision depend on the numerical environment.}
\label{tab:analytic-validation}
\scriptsize
\setlength{\tabcolsep}{2.5pt}
\resizebox{\linewidth}{!}{
\begin{tabular}{lrrrrrr}
\toprule
Propagator & $E_{L^2}$ & $E_{\rm sg}$ & $R_{\rm gen}$ & $E_{\rm ref}$ & $E_{\rm bal}$ & $E_{\rm pert}$ \\
\midrule
Correct FFT & $1.07\times10^{-16}$ & $7.83\times10^{-17}$ & $1.000$ & $0$ & $0$ & $5.50\times10^{-13}$ \\
Frozen map & $0.073$ & $0$ & $0$ & $8.86\times10^{-3}$ & $0$ & $3.825$ \\
Clock defect & $8.96\times10^{-3}$ & $4.43\times10^{-3}$ & $0.997$ & $9.04\times10^{-4}$ & $0$ & $0.245$ \\
Excess diffusion & $0.055$ & $7.18\times10^{-17}$ & $2.099$ & $0.010$ & $0$ & $0.986$ \\
Source leak & $0.079$ & $8.63\times10^{-17}$ & $0.704$ & $0.164$ & $0.038$ & $6.17\times10^{-13}$ \\
\bottomrule
\end{tabular}

}
\end{table}

The FFT map agrees with the closed form near machine precision (Table~\ref{tab:analytic-validation}). The frozen map has zero composition and generator discrepancies but nonzero state/response errors. Excess diffusion preserves composition and dissipates energy, yet gives nonzero reference-energy error. The uniform source preserves composition but violates mass balance. The clock defect breaks composition despite a generator ratio near one.

\subsection{Balance quadrature and failure regressions}
Section~\ref{sec:manufactured} verifies balance weights, flux signs, sources, point faces, and quadrature refinement.

Regressions check nonfinite predictions, failed callbacks, invalid channels, discrete horizons, saved-prediction applicability, reader/model shapes, split disjointness, resume invalidation, and incomplete/obsolete table evidence. A float64 test retains a relative signal of order \(10^{-8}\); generator ratios are excluded from monotone rankings.

\subsection{Installation and reproduction}
\label{sec:installation}
\ref{app:reproduction} gives CPU-only commands. The demonstration exports CSV, strict JSON, and \texttt{diagnostics.tex}, retaining all three horizons and sample counts. Its manifest records version/source hash, dependencies, dtype, device, grid, PDE parameters, and timing. Sixteen checks use cross-platform tolerances and a nonzero failure exit. CI builds both distributions and tests the installed wheel outside the checkout, including training/resume on generated data. Section~\ref{sec:cost} reports environments, test counts, and CPU/CUDA costs.

During the revision stage, OpenAI Codex, using GPT-6.1 Sol, GPT-6 Sol, and GPT-6 Astra (OpenAI), assisted in developing the diagnostic software, regression tests, and experiment and analysis scripts. Numerical figures are rendered from archived artifacts with Matplotlib; the architecture flowchart is drawn from TikZ source developed with Codex assistance.

\section{Numerical sensitivity and independent verification}
\label{sec:sensitivity}
\subsection{Exact heat equation and finite-horizon baseline}
\label{sec:heat_sensitivity}
For periodic heat flow \(u_t=\nu u_{xx}\) on the unit interval, a mode \(v_k\) evolves as \(\Phi_hv_k=e^{-\lambda_kh}v_k\), \(\lambda_k=\nu(2\pi k)^2\). Substitution into the implemented discrepancy at \((h,2h)\) gives, when the denominator dominates epsilon,
\begin{equation}
 E_{\rm gen}^{\rm exact}(h,2h)
 =\frac{1-e^{-\lambda_k h}}{3+e^{-\lambda_k h}}
 =\frac{\lambda_k h}{4}+O((\lambda_kh)^2).
\label{eq:heat_generator}
\end{equation}
Verification retains epsilon and the state norm. The discrepancy increases with timestep and frequency even for exact evolution. For sufficiently regular states, \((\Phi_h-I)u/h=Au+(h/2)A^2u+O(h^2)\); controlling \(A^2u\) explains the regularity requirement.

The sweep uses \(\nu=0.01\), \(N=32,64,128,256\), modes \(k=1,2,4,8,12\), float32/float64, and 25 logarithmically spaced horizons \(10^{-8}\leq h\leq10^{-0.7}\), paired with \(2h\). Continuum evolution and exact centered-Laplacian evolution, with rate \(\lambda_{k,N}=4\nu N^2\sin^2(\pi k/N)\), separate spatial bias from finite-horizon bias without time-integrator error. All modes are below Nyquist. The 2,000 single-mode cases are supplemented by 300 fixed 12-mode cases with amplitudes \(k^{-p}\), \(p=1,2,4\), varying spectral content at fixed maximum frequency.

\begin{figure}[htbp]
\centering
\includegraphics[width=\linewidth]{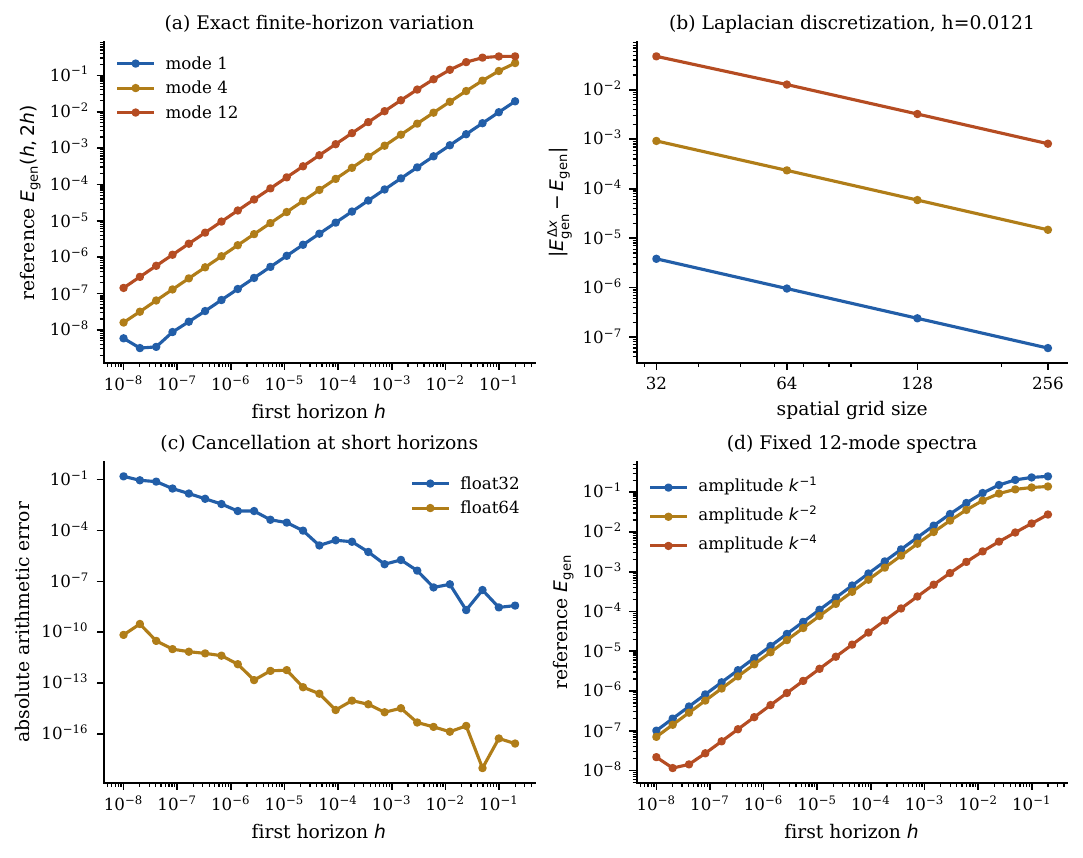}
\caption{Generator sensitivity from archived CSV output. (a) Exact continuum variation depends on frequency and horizon. (b) Centered-Laplacian spatial error decreases under refinement. (c) Cancellation makes very short float32 horizons unreliable. (d) Changing spectral content changes the reference baseline at the same timestep. Panels (a), (c), and (d) use \(N=64\); panel (b) uses \(h=0.0121153\). Zero arithmetic errors are plotted at \(10^{-18}\) and retained as zero in the CSV.}
\label{fig:generator_sensitivity}
\end{figure}
\FloatBarrier

\noindent\begin{minipage}{\linewidth}
For float64 and \(h\geq10^{-4}\), the epsilon-aware formula error is below \(3.2\times10^{-13}\); small-horizon halving gives factor \(1.99999\). At \(k=1,h=0.0121153\), spatial-refinement factors \(3.9962\), \(3.9990\), and \(3.9998\) indicate second-order error. The fixed physical 12-mode spectrum changes discrepancy by less than \(4.3\times10^{-11}\) across grids at this horizon, including floating-point and epsilon effects.
\end{minipage}

\subsection{Reference normalization, precision, and interpretation}
\label{sec:ratio_interpretation}
\begin{figure}[htbp]
\centering
\includegraphics[width=\linewidth]{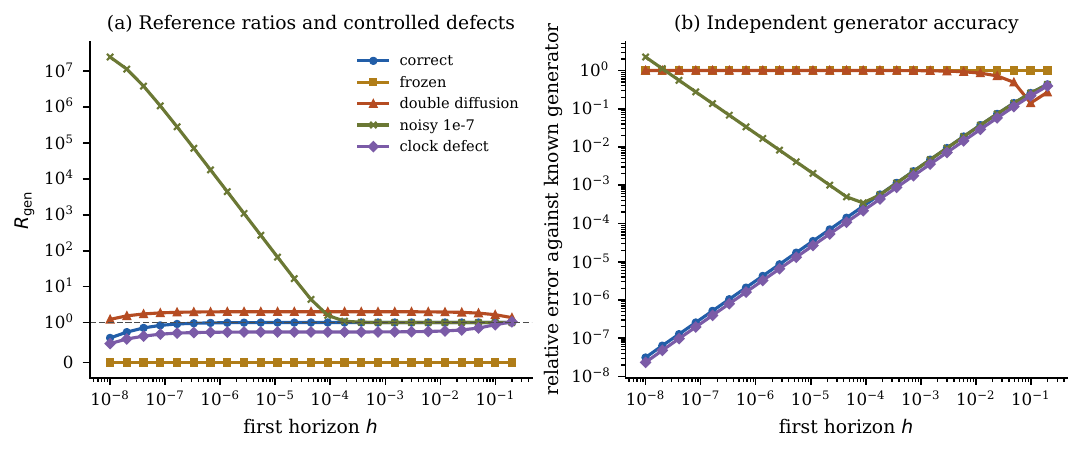}
\caption{Reference normalization is descriptive. The frozen map has \(R_{\rm gen}=0\) while its known-generator error is one. For the correct map the ratio approaches one when the reference scale dominates epsilon; its finite-horizon generator bias remains. Endpoint noise is amplified at short horizons. Ratios are omitted when the reference discrepancy is no greater than epsilon. The ratio axis is linear near zero and logarithmic outside that region.}
\label{fig:ratios}
\end{figure}

Figure~\ref{fig:ratios} audits mode \(k=4\) with freezing, doubled diffusivity, a clock defect, and endpoint noise of amplitude \(10^{-7}\). The 125 cases retain raw discrepancies, \(R_{\rm gen}\), and error against the known generator \(Au=-\lambda_ku\), separating finite-horizon variation from generator accuracy.

A ratio above/below one means more/less finite-horizon variation than the reference. Neither direction ranks dynamics accuracy: the frozen map gives zero. The table builder excludes this ratio from monotone rankings.

Comparisons must match horizons, units, grid, preprocessing, precision, and reference solver. Epsilon biases small-reference ratios below one; short-horizon division amplifies cancellation and endpoint noise, especially in float32. Reports pair the ratio with state error and raw discrepancies. Nearby-horizon/resolution sweeps locate these numerical effects.

\subsection{Perturbation and scaling verification}
\label{sec:response_verification}
The response sweep evaluates heat flow from \(1+0.2\sin(2\pi x)\) and perturbation \(\delta\cos(6\pi x)\), for \(\delta=10^{-2},10^{-3},10^{-4},10^{-5},10^{-6}\), horizons \(0.05,0.1,0.2\), and scales \(0.5,1.25,2\). All 135 cases independently supply base/perturbed reference and prediction states.

\begin{figure}[htbp]
\centering
\includegraphics[width=\linewidth]{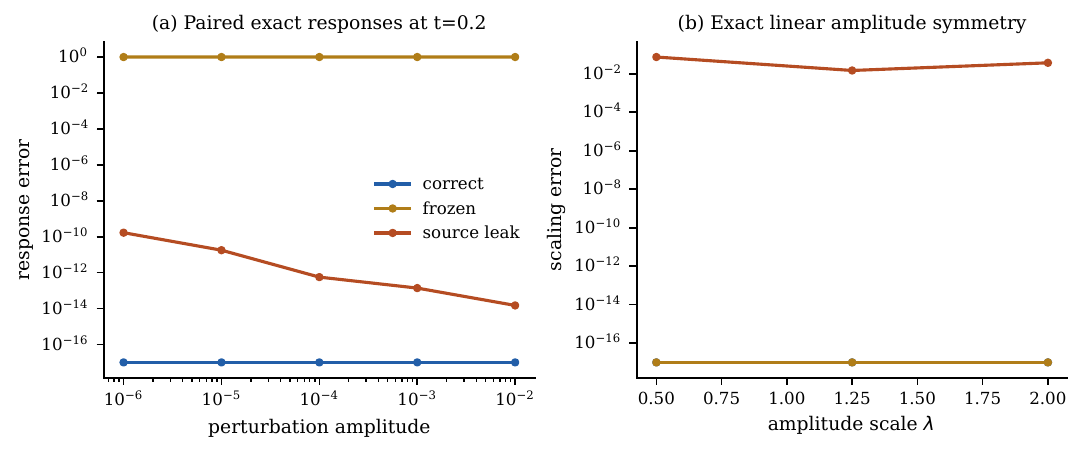}
\caption{Positive and deliberately defective response/scaling verification at \(t=0.2\). Correct heat flow passes both. A frozen map fails perturbation response while satisfying amplitude scaling; a uniform source leak cancels in a paired response but breaks scaling. Exact zeros are displayed at \(10^{-17}\). Panel (a) fixes \(\lambda=1.25\); panel (b) fixes \(\delta=10^{-3}\).}
\label{fig:response_scaling}
\end{figure}

Correct heat flow gives zero errors in float64. The frozen map gives response errors \(0.194\)--\(1.035\) and zero scaling error. The input-independent source \(0.4t\) gives scaling errors \(0.00389\)--\(0.07354\) but response errors below \(1.7\times10^{-10}\), since it cancels from paired differences. Amplitude symmetry follows from linear homogeneity.

\subsection{Manufactured physical residuals}
\label{sec:manufactured}
For \(q_t+\nabla\cdot F=S\), fields \(F=(x_1,\ldots,x_d)\), \(S=2\), \(q(t)=10+(2-d)t\) cancel below \(10^{-14}\) for \(d=1,2,3\), on full domains and subvolumes. With \(F=x^3\), \(q(t,x)=10-3tx^2\), \(S=0\), spatial doubling from 16 to 128 cells gives reduction factors \(3.803\), \(3.904\), \(3.953\). For \(q_t=q\), \(q=e^t\), \(F=0\), horizon halving from \(0.1\) to \(0.0125\) gives \(7.804\), \(7.901\), \(7.950\): third-order reduction of the single-interval trapezoidal balance residual.

Periodic \(B=(\sin y,\sin x)\) gives zero centered divergence; \(B=(\sin x,0)\) gives normalized residual one up to epsilon, checking both stencil and normalization.

\section{Controlled study of learned propagators}
\label{sec:learned}
\subsection{Equations, independent initial states, and propagation scheme}
\label{sec:controlled_propagation}
The confirmatory study uses periodic heat, scalar advection, and viscous Burgers flow on the unit interval:
\[
 u_t=\nu u_{xx},\qquad u_t+cu_x=0,\qquad
 u_t+uu_x=\nu u_{xx}.
\]
Heat uses \(\nu=0.01\), advection \(c=0.3\), and Burgers \(\nu=0.02\), with maximum horizon \(T=0.2\). Each model directly predicts from \(N=64\) samples and horizon \(\tau=t/T\), imposing identity at \(t=0\).

Each equation has disjoint sets of 512 training, 128 validation, and 256 test coefficient identities, shared across seeds \(0,\ldots,4\); initialization and minibatches vary. Heat/advection states have mean \(0.25\), six Fourier modes with amplitude decay \(k^{-3/2}\), and fluctuation RMS uniform in \([0.25,0.5]\). Independent references use modal damping or translation.

For Burgers, four random modes define a potential
\[
 \phi_0(x)=1+\sum_{k=1}^4[a_k\cos(2\pi kx)+b_k\sin(2\pi kx)].
\]
Rescaling to \(\sum_k(|a_k|+|b_k|)=0.6\) preserves positivity. Heat evolution of \(\phi\) at viscosity \(0.02\) gives the Galilean-transformed Cole--Hopf reference:
\[
 u(x,t)=0.25-2\nu
 \frac{\phi_x(x-0.25t,t)}{\phi(x-0.25t,t)}.
\]
Tests compare its finite-time derivative with the Burgers right-hand side on a finer grid. Each scalar field is a complete evolution state.

\subsection{Matched architecture and training interventions}
\label{sec:pino}
All arms use a periodic FNO with four spectral layers, width 32, 12 Fourier modes, GELU, and pointwise channel mixing. A fixed affine transformation uses training initial states. State/time inputs give output \(u+\tau\sigma f_\theta(u,\tau)\), where \(\sigma\) is their standard deviation.

Per seed, a 1,500-update supervised warmup branches from its validation-selected weights and optimizer state into three 1,000-update continuations: supervised, PINO-style PDE-residual \cite{li2024pino}, and composition training. All share batch size 64, state/time minibatches, architecture, labelled data, and update count. AdamW uses initial rate \(0.002\), cosine decay to one tenth, weight decay \(10^{-6}\), and gradient clipping at one. Added physics/composition calls increase wall time.

The supervised loss averages squared per-sample relative error. The PINO-style loss adds
\[
 0.05\,\frac{\langle|\partial_t\widehat u-\mathcal R(\widehat u)|^2\rangle}
 {\max(\langle|\mathcal R(u_0)|^2\rangle_{\rm train},10^{-6})}.
\]
Physics uses \(N=128\) samples of the same initial conditions, Fourier spatial derivatives, and centered time differences with half-spacing \(10^{-3}T\), collocated within \((0,T)\). Composition training adds ten times the squared relative direct-versus-two-equal-substep difference.

These equations supply evolution residuals directly, motivating PINO-style training rather than VINO's variational/energy formulation \cite{eshaghi2025vino}. The arm uses the common FNO, without reproducing published PINO benchmarks or instance-wise fine-tuning.

\subsection{Selection, uncertainty, and accurate-model contrasts}
\label{sec:matched}
Every 50 updates, checkpoints are scored by validation relative \(L^2\) averaged over \(T/8,T/4,T/2,T\). Each arm retains its minimum; nonsupervised arms also retain the checkpoint closest to the supervised minimum in absolute log-score ratio. Matching uses neither test data nor semigroup error. Eligibility requires both validation errors below \(0.02\) and ratio in \([0.8,1.25]\).

Test audits use all 256 states at four horizons, composition \((t/2,t/2)\), and generator pair \((t/2,t)\) for model and reference. Fluctuation errors subtract the shared test initial-state mean. A paired crossed bootstrap resamples training seeds and shared test-state identities independently, preserving model pairings: 10,000 replicates, 95\% percentile intervals.

\begin{table}[htbp]
\centering
\caption{Confirmatory audit of every arm's best validation checkpoint at \(T=0.2\), summarizing five seeds. State, fluctuation, and semigroup errors are percentages; generator ratios use \((T/2,T)\).}
\label{tab:controlled}
\small
\resizebox{\linewidth}{!}{
\begin{tabular}{llrrrr}
\toprule
Equation & Training & State (\%) & Fluctuation (\%) & Semigroup (\%) & \(R_{\rm gen}\)\\
\midrule
Heat & Supervised & \(0.087\pm0.009\) & \(0.113\pm0.011\) & \(0.090\pm0.011\) & \(0.996\pm0.002\)\\
Heat & PINO-style & \(0.050\pm0.013\) & \(0.065\pm0.017\) & \(0.045\pm0.008\) & \(1.001\pm0.001\)\\
Heat & Composition & \(0.086\pm0.008\) & \(0.112\pm0.012\) & \(0.048\pm0.003\) & \(0.999\pm0.002\)\\
Advection & Supervised & \(0.227\pm0.049\) & \(0.272\pm0.060\) & \(0.231\pm0.057\) & \(0.996\pm0.002\)\\
Advection & PINO-style & \(0.138\pm0.035\) & \(0.166\pm0.043\) & \(0.136\pm0.034\) & \(0.999\pm0.002\)\\
Advection & Composition & \(0.243\pm0.017\) & \(0.293\pm0.021\) & \(0.145\pm0.018\) & \(1.000\pm0.001\)\\
Burgers & Supervised & \(0.116\pm0.013\) & \(0.618\pm0.067\) & \(0.112\pm0.009\) & \(0.999\pm0.001\)\\
Burgers & PINO-style & \(0.100\pm0.009\) & \(0.532\pm0.048\) & \(0.094\pm0.007\) & \(1.000\pm0.001\)\\
Burgers & Composition & \(0.150\pm0.021\) & \(0.796\pm0.113\) & \(0.093\pm0.005\) & \(1.001\pm0.002\)\\
\bottomrule
\end{tabular}

}
\end{table}

\begin{table}[htbp]
\centering
\caption{Matching attempts at \(T\), pooling all five seeds, including ineligible ones. Eligibility counts four-horizon validation matches; displayed errors use held-out endpoints. Intervals are paired crossed-bootstrap 95\% intervals. The semigroup factor is supervised/composition error.}
\label{tab:matched}
\small
\resizebox{\linewidth}{!}{
\begin{tabular}{lrrrrr}
\toprule
Equation & Eligible & Supervised \(L^2\) (\%) & Composition \(L^2\) (\%) & SG reduction & Gap (pp)\\
\midrule
Heat & 4/5 & \(0.087\) & \(0.087\) & \(1.86\;[1.65,2.07]\) & \(0.041\;[0.031,0.052]\)\\
Advection & 5/5 & \(0.227\) & \(0.260\) & \(1.48\;[1.34,1.62]\) & \(0.075\;[0.048,0.105]\)\\
Burgers & 0/5 & \(0.116\) & \(0.150\) & \(1.20\;[1.14,1.27]\) & \(0.019\;[0.013,0.025]\)\\
\bottomrule
\end{tabular}

}
\end{table}
\FloatBarrier

For advection, all five seeds meet the validation matching criterion. At the held-out terminal horizon, supervised and matched composition models have state errors 0.227\% and 0.260\%, respectively. Their semigroup errors are 0.231\% and 0.156\%, a reduction factor 1.48 (95\% interval [1.34,1.62]). Heat has 4/5 eligible seeds; Burgers has 0/5. The Burgers composition arm increases prediction error as it reduces its penalized composition residual. Table~\ref{tab:controlled} reports the best-validation checkpoints and fluctuation errors for every arm.

\subsection{Composed prediction and supplementary out-of-horizon rollout}
\label{sec:rollout}
This supplementary audit reuses all best-validation and matched checkpoints without reselection or retraining. Composed error compares \(\widehat\Phi_{T/2}\circ\widehat\Phi_{T/2}\) with the exact reference at \(T\). Two/four calls of duration \(T\), or four/eight of \(T/2\), reach \(2T,4T\); every query stays within \(0<\tau\leq1\). References use the actual later times. Excess errors subtract one-call error at \(T\), comparing different reference times.

The same crossed bootstrap retains all five seeds and arms. Tables retain duplicate selections; correlations count identical weights once per equation/seed. Pooled correlations fix equation strata and independently resample their seeds and test-state identities.

\begin{landscape}
\noindent\begin{minipage}{\linewidth}
\makeatletter\def\@captype{table}\makeatother
\centering
\caption{Supplementary state errors (\%, paired crossed-bootstrap 95\% intervals). Best minimizes validation error; match attempts validation-only accuracy matching. All five seeds and unsuccessful matches are retained. Direct/composed errors use \(T\); chains use duration-\(T\) calls. Half-step chains, excess errors, and partition gaps appear in \ref{app:extended}.}
\label{tab:rollout}
\footnotesize
\resizebox{\linewidth}{!}{
\begin{tabular}{llrrrr}
\toprule
Equation & Training/selection & Direct \(T\) & Composed \(T\) & Chain \(2T\) & Chain \(4T\)\\
\midrule
Heat & Supervised/best & \(0.087\,[0.076,0.099]\) & \(0.048\,[0.043,0.053]\) & \(0.072\,[0.060,0.085]\) & \(0.093\,[0.065,0.124]\)\\
Heat & PINO-style/best & \(0.050\,[0.040,0.062]\) & \(0.043\,[0.033,0.057]\) & \(0.056\,[0.048,0.066]\) & \(0.087\,[0.075,0.102]\)\\
Heat & PINO-style/match & \(0.083\,[0.070,0.096]\) & \(0.060\,[0.052,0.069]\) & \(0.099\,[0.082,0.117]\) & \(0.155\,[0.132,0.179]\)\\
Heat & Composition/best & \(0.086\,[0.076,0.096]\) & \(0.075\,[0.065,0.086]\) & \(0.099\,[0.086,0.113]\) & \(0.136\,[0.106,0.170]\)\\
Heat & Composition/match & \(0.087\,[0.078,0.096]\) & \(0.076\,[0.066,0.087]\) & \(0.101\,[0.090,0.114]\) & \(0.137\,[0.110,0.171]\)\\
Advection & Supervised/best & \(0.227\,[0.185,0.272]\) & \(0.171\,[0.145,0.197]\) & \(0.443\,[0.360,0.532]\) & \(0.890\,[0.725,1.069]\)\\
Advection & PINO-style/best & \(0.138\,[0.110,0.171]\) & \(0.154\,[0.121,0.195]\) & \(0.259\,[0.208,0.322]\) & \(0.511\,[0.410,0.634]\)\\
Advection & PINO-style/match & \(0.209\,[0.172,0.250]\) & \(0.243\,[0.196,0.294]\) & \(0.401\,[0.332,0.478]\) & \(0.793\,[0.659,0.941]\)\\
Advection & Composition/best & \(0.243\,[0.218,0.272]\) & \(0.218\,[0.191,0.253]\) & \(0.475\,[0.427,0.536]\) & \(0.951\,[0.851,1.079]\)\\
Advection & Composition/match & \(0.260\,[0.226,0.298]\) & \(0.233\,[0.196,0.276]\) & \(0.512\,[0.444,0.589]\) & \(1.025\,[0.887,1.183]\)\\
Burgers & Supervised/best & \(0.116\,[0.104,0.133]\) & \(0.066\,[0.061,0.071]\) & \(0.112\,[0.096,0.136]\) & \(0.155\,[0.126,0.182]\)\\
Burgers & PINO-style/best & \(0.100\,[0.089,0.112]\) & \(0.086\,[0.077,0.098]\) & \(0.118\,[0.098,0.141]\) & \(0.254\,[0.178,0.347]\)\\
Burgers & PINO-style/match & \(0.110\,[0.099,0.124]\) & \(0.094\,[0.083,0.104]\) & \(0.130\,[0.107,0.164]\) & \(0.271\,[0.184,0.388]\)\\
Burgers & Composition/best & \(0.150\,[0.130,0.171]\) & \(0.127\,[0.102,0.158]\) & \(0.154\,[0.119,0.197]\) & \(0.257\,[0.127,0.421]\)\\
Burgers & Composition/match & \(0.150\,[0.130,0.171]\) & \(0.127\,[0.102,0.158]\) & \(0.154\,[0.119,0.197]\) & \(0.257\,[0.127,0.421]\)\\
\bottomrule
\end{tabular}

}
\end{minipage}
\end{landscape}

For advection, the supervised and matched-composition models have composed-at-\(T\) state errors 0.171\% and 0.233\%, respectively. Their full-step \(4T\) rollout errors are 0.890\% and 1.025\%; the supervised/composition ratio is \(0.868\,[0.792,0.943]\). The composition penalty lowers disagreement between time partitions while increasing both composed-state and longer-rollout error. The supervised arm gives the more accurate forecast.

Spearman correlations between semigroup error and full-step \(4T\) rollout error are Heat: \(\rho=0.092\,[-0.245,0.335]\); Advection: \(\rho=0.458\,[0.192,0.632]\); Burgers: \(\rho=0.063\,[-0.510,0.490]\); Pooled equations: \(\rho=0.840\,[0.745,0.882]\). The heat and Burgers intervals cross zero. The larger pooled coefficient combines within-equation association with differences between equations. These descriptive correlations summarize the audited checkpoints; prediction for unseen models would require a separate test. All \(2T\), \(4T\), and half-step coefficients are archived.

\subsection{Identical-endpoint construction check}
\label{sec:matched_clocks}
For each supervised heat/advection FNO, clocks \(c_g(t)=gt\) and \(q_g(t)=t[1+(g-1)t/t_*]\), \(t_*=T/2\), query the same network. Endpoints at \(t_*\) are identical; half-time queries differ. Validation state error calibrates \(g\in[1,2]\) to 1\% using 24 bisections. All queries stay within the trained horizon range; all five seeds and 256 test states are retained.

\begin{table}[htbp]
\centering
\caption{Designed-clock check at \(t_*=0.1\), separate from training-arm comparisons. Paired endpoints are bitwise identical. SG factors compare quadratic/constant clocks, using the same crossed bootstrap.}
\label{tab:matched_clocks}
\small
\resizebox{\linewidth}{!}{
\begin{tabular}{lrrrr}
\toprule
Equation & Identical state (\%) & Constant SG (\%) & Quadratic SG (\%) & SG factor [95\% interval]\\
\midrule
Heat & \(1.011\) & \(0.047\) & \(0.513\) & \(10.88\;[9.15,12.82]\)\\
Advection & \(1.004\) & \(0.157\) & \(0.544\) & \(3.46\;[3.08,3.91]\)\\
\bottomrule
\end{tabular}

}
\end{table}

Both clocks alter the target dynamics, including the linearly rescaled clock satisfying clock composition.

\subsection{Fixed-step autoregressive control}
\label{sec:autoregressive_control}
Each supervised model is fixed at \(T/8\) and repeated eight times to \(T\). Table~\ref{tab:autoregressive} checks reference accuracy and the construction identity from Section~\ref{sec:semigroup}.

\begin{table}[htbp]
\centering
\caption{Fixed-step control using the supervised model at \(T/8\). State error compares the eight-call endpoint with the reference at \(T\); the composition column checks the algebraic identity.}
\label{tab:autoregressive}
\small

\begin{tabular}{lrr}
\toprule
Equation & State error (\%) & Semigroup residual\\
\midrule
Heat & \(0.133\pm0.042\) & \(0.0e+00\)\\
Advection & \(0.561\pm0.158\) & \(0.0e+00\)\\
Burgers & \(0.158\pm0.015\) & \(0.0e+00\)\\
\bottomrule
\end{tabular}

\end{table}

\section{External benchmark application}
\label{sec:external_results}
The five tasks use the contracts in Section~\ref{sec:external}. Before fitting, Well selection takes the lexicographically first filename common to official train/validation/test directories, fixing active-matter \(L=10,\zeta=11,\alpha=-2\) and MHD \({\rm Ma}=0.7,{\rm Ms}=0.5\). Archived manifests retain file lists, shapes, spacing, fields, trajectory identities, and SHA-256 hashes.

FNO, branch/trunk DeepONet, U-Net, and ResNet each use one direct call at \(\tau=k/4\). Composition adds calls at \(\lfloor k/2\rfloor/4\) and \(\lceil k/2\rceil/4\), restarting from the decoded intermediate state. Model and reference generator pairs use one stored step and the requested later step.

Seeds \(0,1,2\) use up to 1,024 training pairs, 256 validation pairs, and 128 test windows each, except Active2D (77) and MHD3D (96). Exact selections and model counts are archived. Width is 24, with eight Fourier modes. Standard/short budgets use 30/two epochs with identical initialization and minibatches; checkpoints minimize validation normalized MSE. DeepONet runs only in 2D because of memory cost. Spatial upper limits are 32 (2D) and 16 (3D), preserving uniform periodic seams.

Oversmoothing uses Section~\ref{sec:smoothing}. Windows vary by seed and may overlap; some Well files contain one independent test trajectory. Seed spread is descriptive.

\begin{table}[htbp]
\centering
\caption{FNO audit at horizon two, averaged over three seeds. Functionals include the active squared norm and partial MHD energy. Dashes mean unavailable. Full coverage and constraint components appear in \ref{app:extended}.}
\label{tab:external}
\footnotesize
\resizebox{\linewidth}{!}{
\begin{tabular}{llrrrrrr}
\toprule
Task/model & Budget/filter & State (\%) & SG (\%) & Functional & Balance & \(R_{\rm gen}\) & Scaling\\
\midrule
ADV2D & Standard & \(0.858\pm0.1\) & \(1.17\pm0.11\) & \(0.00613\pm0.00059\) & --- & \(1\pm0.0055\) & \(0.0202\pm0.0026\)\\
ADV2D & Short & \(5.53\pm0.26\) & \(7.72\pm0.41\) & \(0.0574\pm0.015\) & --- & \(1.15\pm0.054\) & \(0.0876\pm0.0088\)\\
ADV2D & Smoothed & \(40.5\pm0.099\) & \(29.5\pm0.19\) & \(0.537\pm0.0022\) & --- & \(1.82\pm0.0046\) & \(0.0178\pm0.0024\)\\
SW2D & Standard & \(0.261\pm0.02\) & \(0.214\pm0.015\) & --- & --- & \(1.44\pm0.088\) & ---\\
SW2D & Short & \(0.626\pm0.013\) & \(1.02\pm0.13\) & --- & --- & \(3.69\pm0.3\) & ---\\
SW2D & Smoothed & \(3.32\pm0.047\) & \(1.12\pm0.052\) & --- & --- & \(4.13\pm0.058\) & ---\\
Active2D & Standard & \(22.8\pm0.87\) & \(10.7\pm0.72\) & \(0.0333\pm0.0061\) & --- & \(1.06\pm0.053\) & ---\\
Active2D & Short & \(25.3\pm0.41\) & \(18.7\pm0.24\) & \(0.0639\pm0.015\) & --- & \(1.38\pm0.021\) & ---\\
Active2D & Smoothed & \(36.4\pm0.28\) & \(13.2\pm0.32\) & \(0.224\pm0.0021\) & --- & \(1.42\pm0.0018\) & ---\\
CNS3D & Standard & \(13.6\pm1.3\) & \(8.22\pm1\) & \(0.0561\pm0.021\) & \(0.0105\pm0.0048\) & \(0.636\pm0.027\) & ---\\
CNS3D & Short & \(21.3\pm2.1\) & \(15.4\pm2.5\) & \(0.176\pm0.021\) & \(0.0196\pm0.0035\) & \(0.609\pm0.052\) & ---\\
CNS3D & Smoothed & \(9.43\pm1\) & \(7.01\pm3.4\) & \(0.0561\pm0.021\) & \(0.0105\pm0.0048\) & \(0.63\pm0.027\) & ---\\
MHD3D & Standard & \(27.7\pm0.23\) & \(12.8\pm2.1\) & \(0.0747\pm0.017\) & \(0.0016\pm0.00088\) & \(1.1\pm0.066\) & ---\\
MHD3D & Short & \(27.7\pm0.17\) & \(15.1\pm1.9\) & \(0.101\pm0.026\) & \(0.00385\pm0.0027\) & \(1.21\pm0.051\) & ---\\
MHD3D & Smoothed & \(43\pm0.0064\) & \(12.7\pm1.8\) & \(0.381\pm0.014\) & \(0.0016\pm0.00088\) & \(1.25\pm0.0033\) & ---\\
\bottomrule
\end{tabular}

}
\end{table}

The application contains 531 per-seed/horizon records: 486 learned-model/variant records and 45 reference records. All declared task, architecture, variant, seed, and horizon combinations are present, with finite applicable values. State errors are appreciably larger than in the controlled study. \ref{app:extended} reports every supported architecture at horizon two; the CSV retains horizons one, two, and four. Reference generator, balance, and constraint residuals provide baselines for the physical comparisons.

ADV2D scaling uses \(s=1.25\) (Tables~\ref{tab:external},~\ref{tab:external_all}); other tasks have no configured amplitude symmetry.

For CNS3D, reference internal energy contributes 99.82466\% of the retained-grid total. All 1,152 saved-checkpoint smoothing pairs change the predicted state; the FNO reference-functional errors are 0.0561234 (standard) and 0.0560513 (smoothed). The pressure contribution is linear and its periodic spatial mean is preserved by filtering; the small kinetic contribution changes too little to alter the displayed total-functional precision. Full-domain mass balance is likewise insensitive to this mass-preserving filter.

The primary balance uses the periodic full domain. A checkpoint audit also evaluates its first octant---half of each grid axis---for every CNS3D/MHD3D model/variant at horizon two (Table~\ref{tab:external_subvolumes}). Octant kernels use float64 on float32 predictions; primary records use float32. Reference baselines retain reconstruction and time-quadrature residuals.

Shallow-water height omits momentum. Its composition score can reflect missing state information as well as surrogate error.

\section{Computational cost and release verification}
\label{sec:cost}
Endpoint state, functional, positivity, and whole-domain balance kernels cost work linear in state size. Divergence adds \(O(d^2)\) derivatives for gradient normalization. Panel latency includes validity checks, tensor-to-array transfer, and aggregation.

For \(H\) horizons, direct endpoints require \(H\) calls. Per horizon, composition adds two, perturbation adds one, and scaling adds two to allow time rescaling. Generators reuse endpoints. A full two-horizon audit makes 12 calls versus two for endpoints alone; saved-array audits make none.

\begin{table}[htbp]
\centering
\caption{Representative synchronized median latencies in milliseconds, batch size four. Two endpoint predictions use two callable evaluations; the full configured two-horizon audit uses 12. The saved endpoint panel makes no model calls. CUDA memory is additional allocator peak for the full panel, in MiB. All grids, precisions, quartiles, CPU peak RSS, and raw 31-repeat summaries are archived.}
\label{tab:overhead}
\small
\resizebox{\linewidth}{!}{
\begin{tabular}{lllrrrrr}
\toprule
Device & Grid & Precision & State only & Saved panel & Two predictions & Full panel & CUDA MiB\\
\midrule
CPU & \(64^2\) & float32 & 0.010 & 0.550 & 0.159 & 3.579 & ---\\
CPU & \(64^2\) & float64 & 0.010 & 0.533 & 0.118 & 3.375 & ---\\
CPU & \(32^3\) & float32 & 0.026 & 2.668 & 1.680 & 18.005 & ---\\
CPU & \(32^3\) & float64 & 0.044 & 2.573 & 1.042 & 25.130 & ---\\
CUDA & \(64^2\) & float32 & 0.024 & 0.838 & 0.062 & 3.679 & 1.19\\
CUDA & \(64^2\) & float64 & 0.029 & 0.845 & 0.111 & 3.748 & 2.38\\
CUDA & \(32^3\) & float32 & 0.023 & 0.854 & 0.068 & 3.833 & 9.50\\
CUDA & \(32^3\) & float64 & 0.036 & 0.887 & 0.124 & 4.069 & 19.00\\
\bottomrule
\end{tabular}}
\end{table}

\FloatBarrier

Measurements use a periodic advection--diffusion FFT callable with state, dissipative-functional, mass-balance, positivity, response, and scaling contracts: five warmups, 31 repetitions, batch size four, float32/float64, 2D grids \(32,64,128\), and 3D grids \(16,32,64\). Hardware is an Intel Core Ultra 9 285K and NVIDIA RTX PRO 6000 Blackwell Workstation Edition. CPU uses one PyTorch thread; CUDA timing is synchronized. CPU memory is total peak RSS, including Python/PyTorch; CUDA memory is additional allocator peak after warmup. Other propagators' costs depend on channels, callbacks, inference, and transfers.

All 106 tests passed against the source and against the installed version 0.3.3 wheel outside the checkout in both dependency environments. The minimum CPU environment uses PyTorch 2.0.1+cpu, NumPy 1.23.5, pandas 1.5.3, h5py 3.8.0; the current environment uses PyTorch 2.11.0.dev20260106+cu128, NumPy 1.24.4, pandas 2.3.3, h5py 3.14.0, with Python 3.10.19. Both installations passed the analytic self-check, 16 referee checks, nine sensitivity checks, public examples, the manufactured balance/constraint script, and a four-model CPU training/resume check. Resume preserved all completed checkpoint bytes.

Tests cover the regressions in Section~\ref{sec:analytic}, cached/original data equality, and checkpoint preservation. Distributions and numerical artifacts carry \mbox{SHA-256} identifiers.

\section{Limitations and future work}
\label{sec:limitations}
Nonuniform and unstructured grids require new integration weights and face reconstruction. Solver-native fluxes and divergence operators would better match each reference discretization. The controlled comparison covers scalar equations at fixed coefficients; the coarse external grids and single-trajectory Well selections do not quantify uncertainty across physical regimes. Forecast associations across independently developed models remain untested.

\section{Conclusions}
\label{sec:conclusions}
PDECheck combines callable temporal audits with equation-specific physical residuals and verified reporting. The heat sweep separates finite-horizon generator variation from dynamics error. In the matched advection comparison, composition regularization lowers semigroup error but increases composed-state and longer-rollout error. Reporting both partition agreement and reference accuracy makes this distinction visible.

\section*{Declaration of competing interest}
The author received hardware support from Dell Technologies through the Dell Pro Precision Ambassador Program. The author declares no other relevant financial or non-financial interests.

\section*{Funding}
Computational resources used in this work were supported in part by hardware provided by Dell Technologies through the Dell Pro Precision Ambassador Program. Dell Technologies had no role in the study design, experiments, analysis, interpretation of results, or preparation of the manuscript.

\section*{Data availability}
The referee demonstration generates analytic data. Third-party PDEBench, APEBench, and The Well datasets retain their original terms and are not redistributed; file selections and hashes identify the inputs. Reproducibility artifacts include controlled initial coefficients, references, diagnostic records, and checkpoints.

\section*{Code availability}
PDECheck \href{\SoftwareReleaseURL}{version~\PackageVersion} provides source and wheels at \url{https://github.com/lennonshikhman/diagnostics_for_physics}. The distribution is \texttt{pdecheck}; imports use \texttt{diagnostic\_suite}. Tests, examples, table/figure generators, and checkpoint audits are included. \ref{app:reproduction} describes provenance.

\section*{Author contributions}
Lennon J. Shikhman conceived the study, developed the diagnostic framework and software design, implemented the experiments, analyzed the results, and wrote the manuscript.

\appendix

\section{Exact physical contracts and propagation capabilities}
\label{app:contracts}
Table~\ref{tab:physical_contracts} specifies task channels and physical contracts.

\begin{table}[htbp]
\centering
\caption{Constraint and balance contracts, with zero-based channel indices. Components are reported separately.}
\label{tab:physical_contracts}
\footnotesize
\setlength{\tabcolsep}{4pt}
\begin{tabular}{>{\raggedright\arraybackslash}p{0.13\linewidth}>{\raggedright\arraybackslash}p{0.25\linewidth}>{\raggedright\arraybackslash}p{0.28\linewidth}>{\raggedright\arraybackslash}p{0.24\linewidth}}
\toprule
Task & State channels & Constraint components & Balance and boundary\\
\midrule
ADV2D & Scalar \(q\): 0 & None; positivity not required & Global drift; periodic; no flux coefficients\\
SW2D & Height \(h\): 0 & \(E_{\rm pos}(h)\) & Global drift; file grid, nonperiodic edges; no velocity\\
Active2D & Concentration; \(v_x,v_y\); tensors \(D,E\), each \(xx,xy,yx,yy\) & None configured & No balance; periodic file grid\\
CNS3D & \(\rho:0,p:1,v_x:2,v_y:3,v_z:4\) & \(E_{\rm pos}(\rho)\), \(E_{\rm pos}(p)\); no velocity-divergence constraint & \(q=\rho,F=\rho v,S=0\); periodic\\
MHD3D & \(\rho:0,v_x:1,v_y:2,v_z:3,B_x:4,B_y:5,B_z:6\) & \(E_{\rm pos}(\rho)\), \(E_{\rm div}(B)\) & \(q=\rho,F=\rho v,S=0\); periodic\\
\bottomrule
\end{tabular}
\end{table}

For periodic derivatives,
\[
 D_i v_{\boldsymbol j}
 =\frac{v_{\boldsymbol j+\boldsymbol e_i}-v_{\boldsymbol j-\boldsymbol e_i}}{2\Delta x_i},
\]
with wraparound. Divergence is \(\|\sum_iD_iv_i\|_2\); normalization includes all vector components:
\[
 \|\nabla_h v\|_F+\varepsilon
 =\left(\sum_{i,j}\|D_iv_j\|_2^2\right)^{1/2}+\varepsilon.
\]
Nonperiodic interiors use this stencil; edges use first-order one-sided differences. Positivity has no derivatives or ghost values.

The generic reference functional is exactly
\[
 J(u)=\frac{1}{2C N_{\rm cell}}\sum_{c,\boldsymbol j}u_{c,\boldsymbol j}^2.
\]
Active matter uses \(J\) only as a reference norm; scalar advection uses its conservative law, and heat/Burgers its dissipative law. CNS/MHD functionals are defined in Section~\ref{sec:external}. All-channel norms depend on representation and nondimensionalization; physical channel weights are not inferred.

For global drift the exact implementation is
\[
 D_{\rm global}=
 \frac{|\sum_{\boldsymbol j}q_{1,\boldsymbol j}-\sum_{\boldsymbol j}q_{0,\boldsymbol j}|}
 {\sum_{\boldsymbol j}|q_{0,\boldsymbol j}|+
  \sum_{\boldsymbol j}|q_{1,\boldsymbol j}|+\varepsilon}.
\]
Uniform cell volumes cancel. Boundary flux and sources belong to \(E_{\rm bal}\), not \(D_{\rm global}\).

\begin{table}[htbp]
\centering
\caption{Propagation schemes and composition applicability.}
\label{tab:propagation}
\small
\begin{tabular}{p{0.26\linewidth}p{0.31\linewidth}p{0.33\linewidth}}
\toprule
Model/interface & Direct horizon map & Meaning of semigroup value\\
\midrule
Controlled FNO, PINO-style, composition arm & One time-conditioned call; \(\tau=t/T\) & Independent direct-versus-two-call comparison\\
External FNO, DeepONet, U-Net, ResNet & One call at \(\tau=k/4\) & Two restarted calls at integer horizons summing to \(k\); unavailable at one stored step\\
Fixed-step control/adapter & Repeated powers \(F^n\) & Equality by construction at integer steps; not independent validation\\
Saved direct endpoints & Supplied endpoint arrays & Unavailable without actual composed endpoint arrays\\
\bottomrule
\end{tabular}
\end{table}

\section{Reproduction and evidence organization}
\label{app:reproduction}
The source provides \path{configs/revision_protocol.json}, fixed-parameter file lists, and scripts for staging, training, analysis, figures, and costs.

The CPU-only numerical checks require no third-party datasets:
\begin{Verbatim}[fontsize=\small]
python -m pip install .
diagnostic-suite --self-check
diagnostic-suite-referee --outdir outputs_referee
diagnostic-suite-sensitivity --outdir outputs_sensitivity
python scripts/verify_balance_and_constraints.py \
    --outdir outputs_balance
python -m pytest
\end{Verbatim}
The full controlled campaign uses three equations and five seeds:
\begin{Verbatim}[fontsize=\small]
diagnostic-suite-controlled --outdir outputs_controlled \
    --warmup-steps 1500 --continuation-steps 1000 \
    --validation-every 50 --composition-weight 10 --resume
\end{Verbatim}
Artifacts retain split coefficients, validation histories, warmup/selected checkpoints, predictions, paired sample metrics, and source/configuration hashes. Resume verifies completed configurations and hashes; incomplete seeds are recomputed. Validation-only pilots used a separate dataset seed and are excluded from test evidence.

Acquire external data from the original projects. Paths/hashes identify exports; manifests flag missing generation metadata. \path{scripts/stage_revision_data.py} copies read-only inputs, verifying source/destination hashes without changing values. \texttt{--file-selection} fixes Well parameters; \texttt{--cache-data} preserves checked pairs' order and values. The evidence bundle includes full commands and runtime details.

Normalization uses training pairs only. Manifests distinguish original training, later checkpoint analysis, and installed-release verification sources; Section~\ref{sec:api} specifies split and reporting checks.

\section{Additional model and physical-component results}
\label{app:extended}
Tables retain every external architecture/variant at horizon two and separate physical components. CSVs retain all three horizons by seed. Spread is sample standard deviation across three seeds, not an independent-trajectory confidence interval.

\begin{landscape}
\scriptsize
\setlength{\tabcolsep}{3pt}

\begin{longtable}{llrrrrrr}
\caption{All supported architectures and variants at horizon two, mean and sample standard deviation across three seeds. State and SG are percentages; other columns are dimensionless. Raw records preserve all three horizons.}\label{tab:external_all}\\
\toprule
Task/model & Budget/filter & State (\%) & SG (\%) & Functional & Balance & \(R_{\rm gen}\) & Scaling\\
\midrule
\endfirsthead
\toprule
Task/model & Budget/filter & State (\%) & SG (\%) & Functional & Balance & \(R_{\rm gen}\) & Scaling\\
\midrule
\endhead
\bottomrule
\endfoot
ADV2D / FNO & Standard & \(0.858\pm0.1\) & \(1.17\pm0.11\) & \(0.00613\pm0.00059\) & --- & \(1\pm0.0055\) & \(0.0202\pm0.0026\)\\
ADV2D / FNO & Short & \(5.53\pm0.26\) & \(7.72\pm0.41\) & \(0.0574\pm0.015\) & --- & \(1.15\pm0.054\) & \(0.0876\pm0.0088\)\\
ADV2D / FNO & Smoothed & \(40.5\pm0.099\) & \(29.5\pm0.19\) & \(0.537\pm0.0022\) & --- & \(1.82\pm0.0046\) & \(0.0178\pm0.0024\)\\
ADV2D / DeepONet & Standard & \(93.1\pm0.067\) & \(32\pm1.9\) & \(0.848\pm0.01\) & --- & \(2.63\pm0.013\) & \(0.112\pm0.014\)\\
ADV2D / DeepONet & Short & \(99.5\pm0.49\) & \(310\pm2.2e+02\) & \(0.983\pm0.016\) & --- & \(2.63\pm0.0086\) & \(0.429\pm0.23\)\\
ADV2D / DeepONet & Smoothed & \(93.2\pm0.13\) & \(32.4\pm0.49\) & \(0.883\pm0.006\) & --- & \(2.63\pm0.013\) & \(0.11\pm0.014\)\\
ADV2D / U-Net & Standard & \(2.07\pm0.11\) & \(2.53\pm0.12\) & \(0.00776\pm0.0072\) & --- & \(0.989\pm0.032\) & \(0.0134\pm0.00059\)\\
ADV2D / U-Net & Short & \(7.85\pm0.26\) & \(9.85\pm0.36\) & \(0.0407\pm0.01\) & --- & \(1.06\pm0.043\) & \(0.0712\pm0.015\)\\
ADV2D / U-Net & Smoothed & \(40.7\pm0.19\) & \(29.8\pm0.25\) & \(0.54\pm0.0029\) & --- & \(1.83\pm0.016\) & \(0.0108\pm0.00078\)\\
ADV2D / ResNet & Standard & \(2.1\pm0.21\) & \(2.44\pm0.39\) & \(0.00355\pm0.001\) & --- & \(1.01\pm0.0039\) & \(0.0101\pm0.00088\)\\
ADV2D / ResNet & Short & \(7.93\pm0.7\) & \(8.5\pm1.4\) & \(0.0156\pm0.0059\) & --- & \(1.09\pm0.062\) & \(0.0358\pm0.0019\)\\
ADV2D / ResNet & Smoothed & \(40.5\pm0.061\) & \(29.5\pm0.11\) & \(0.537\pm0.0015\) & --- & \(1.82\pm0.019\) & \(0.00743\pm0.00051\)\\
SW2D / FNO & Standard & \(0.261\pm0.02\) & \(0.214\pm0.015\) & --- & --- & \(1.44\pm0.088\) & ---\\
SW2D / FNO & Short & \(0.626\pm0.013\) & \(1.02\pm0.13\) & --- & --- & \(3.69\pm0.3\) & ---\\
SW2D / FNO & Smoothed & \(3.32\pm0.047\) & \(1.12\pm0.052\) & --- & --- & \(4.13\pm0.058\) & ---\\
SW2D / DeepONet & Standard & \(2.08\pm0.013\) & \(0.895\pm0.3\) & --- & --- & \(4.17\pm0.19\) & ---\\
SW2D / DeepONet & Short & \(5.22\pm0.28\) & \(2.1\pm0.24\) & --- & --- & \(4.19\pm0.062\) & ---\\
SW2D / DeepONet & Smoothed & \(3.42\pm0.054\) & \(1.33\pm0.29\) & --- & --- & \(4.16\pm0.072\) & ---\\
SW2D / U-Net & Standard & \(0.3\pm0.0024\) & \(0.239\pm0.0031\) & --- & --- & \(1.59\pm0.097\) & ---\\
SW2D / U-Net & Short & \(0.901\pm0.11\) & \(0.916\pm0.068\) & --- & --- & \(3.61\pm0.24\) & ---\\
SW2D / U-Net & Smoothed & \(3.34\pm0.06\) & \(1.25\pm0.0049\) & --- & --- & \(4.12\pm0.043\) & ---\\
SW2D / ResNet & Standard & \(0.215\pm0.017\) & \(0.175\pm0.022\) & --- & --- & \(1.24\pm0.077\) & ---\\
SW2D / ResNet & Short & \(0.563\pm0.086\) & \(0.515\pm0.17\) & --- & --- & \(2.24\pm0.61\) & ---\\
SW2D / ResNet & Smoothed & \(3.34\pm0.063\) & \(1.4\pm0.011\) & --- & --- & \(4.11\pm0.061\) & ---\\
Active2D / FNO & Standard & \(22.8\pm0.87\) & \(10.7\pm0.72\) & \(0.0333\pm0.0061\) & --- & \(1.06\pm0.053\) & ---\\
Active2D / FNO & Short & \(25.3\pm0.41\) & \(18.7\pm0.24\) & \(0.0639\pm0.015\) & --- & \(1.38\pm0.021\) & ---\\
Active2D / FNO & Smoothed & \(36.4\pm0.28\) & \(13.2\pm0.32\) & \(0.224\pm0.0021\) & --- & \(1.42\pm0.0018\) & ---\\
Active2D / DeepONet & Standard & \(63.1\pm0.49\) & \(8.75\pm8.9\) & \(0.381\pm0.0046\) & --- & \(1.51\pm0.00016\) & ---\\
Active2D / DeepONet & Short & \(63.1\pm0.51\) & \(3.75\pm1.7\) & \(0.382\pm0.0044\) & --- & \(1.5\pm0.00059\) & ---\\
Active2D / DeepONet & Smoothed & \(63\pm0.42\) & \(8.53\pm8.4\) & \(0.383\pm0.0038\) & --- & \(1.5\pm0.00011\) & ---\\
Active2D / U-Net & Standard & \(17.3\pm0.28\) & \(8.49\pm0.38\) & \(0.0235\pm0.0013\) & --- & \(0.977\pm0.051\) & ---\\
Active2D / U-Net & Short & \(22.5\pm0.13\) & \(21.7\pm0.84\) & \(0.0655\pm0.0039\) & --- & \(1.46\pm0.017\) & ---\\
Active2D / U-Net & Smoothed & \(35.6\pm0.18\) & \(13.1\pm0.37\) & \(0.224\pm0.0029\) & --- & \(1.43\pm0.015\) & ---\\
Active2D / ResNet & Standard & \(16.9\pm0.26\) & \(7.56\pm0.14\) & \(0.0193\pm0.0064\) & --- & \(0.79\pm0.024\) & ---\\
Active2D / ResNet & Short & \(22.4\pm0.31\) & \(11.3\pm1.1\) & \(0.0404\pm0.014\) & --- & \(1.09\pm0.028\) & ---\\
Active2D / ResNet & Smoothed & \(35.2\pm0.15\) & \(11.7\pm0.47\) & \(0.22\pm0.0056\) & --- & \(1.41\pm0.0078\) & ---\\
CNS3D / FNO & Standard & \(13.6\pm1.3\) & \(8.22\pm1\) & \(0.0561\pm0.021\) & \(0.0105\pm0.0048\) & \(0.636\pm0.027\) & ---\\
CNS3D / FNO & Short & \(21.3\pm2.1\) & \(15.4\pm2.5\) & \(0.176\pm0.021\) & \(0.0196\pm0.0035\) & \(0.609\pm0.052\) & ---\\
CNS3D / FNO & Smoothed & \(9.43\pm1\) & \(7.01\pm3.4\) & \(0.0561\pm0.021\) & \(0.0105\pm0.0048\) & \(0.63\pm0.027\) & ---\\
CNS3D / U-Net & Standard & \(13.9\pm0.71\) & \(17.7\pm5.1\) & \(0.0959\pm0.026\) & \(0.0138\pm0.0083\) & \(0.93\pm0.042\) & ---\\
CNS3D / U-Net & Short & \(25\pm10\) & \(13.7\pm2.7\) & \(0.201\pm0.16\) & \(0.0341\pm0.02\) & \(0.64\pm0.032\) & ---\\
CNS3D / U-Net & Smoothed & \(12\pm0.87\) & \(12.8\pm4.3\) & \(0.0953\pm0.026\) & \(0.0138\pm0.0083\) & \(0.821\pm0.04\) & ---\\
CNS3D / ResNet & Standard & \(11.6\pm1.2\) & \(12.8\pm2.9\) & \(0.0339\pm0.027\) & \(0.00409\pm0.0026\) & \(0.873\pm0.072\) & ---\\
CNS3D / ResNet & Short & \(12.9\pm0.76\) & \(5.54\pm0.99\) & \(0.0329\pm0.022\) & \(0.00356\pm0.002\) & \(0.638\pm0.0062\) & ---\\
CNS3D / ResNet & Smoothed & \(8.29\pm1.2\) & \(7.29\pm2.8\) & \(0.0349\pm0.027\) & \(0.00409\pm0.0026\) & \(0.707\pm0.026\) & ---\\
MHD3D / FNO & Standard & \(27.7\pm0.23\) & \(12.8\pm2.1\) & \(0.0747\pm0.017\) & \(0.0016\pm0.00088\) & \(1.1\pm0.066\) & ---\\
MHD3D / FNO & Short & \(27.7\pm0.17\) & \(15.1\pm1.9\) & \(0.101\pm0.026\) & \(0.00385\pm0.0027\) & \(1.21\pm0.051\) & ---\\
MHD3D / FNO & Smoothed & \(43\pm0.0064\) & \(12.7\pm1.8\) & \(0.381\pm0.014\) & \(0.0016\pm0.00088\) & \(1.25\pm0.0033\) & ---\\
MHD3D / U-Net & Standard & \(24.4\pm0.1\) & \(7.29\pm0.28\) & \(0.086\pm0.0053\) & \(0.00245\pm0.0012\) & \(0.805\pm0.034\) & ---\\
MHD3D / U-Net & Short & \(26.7\pm0.08\) & \(15.3\pm0.35\) & \(0.112\pm0.0094\) & \(0.00311\pm0.00087\) & \(1.26\pm0.0052\) & ---\\
MHD3D / U-Net & Smoothed & \(43\pm0.18\) & \(9.89\pm0.17\) & \(0.392\pm0.0057\) & \(0.00245\pm0.0012\) & \(1.24\pm0.0043\) & ---\\
MHD3D / ResNet & Standard & \(24.1\pm0.041\) & \(6.06\pm0.26\) & \(0.0777\pm0.0035\) & \(0.00209\pm0.0012\) & \(0.626\pm0.015\) & ---\\
MHD3D / ResNet & Short & \(27.1\pm0.37\) & \(8.1\pm0.86\) & \(0.101\pm0.012\) & \(0.00277\pm0.0016\) & \(0.948\pm0.085\) & ---\\
MHD3D / ResNet & Smoothed & \(42.9\pm0.035\) & \(9.88\pm0.55\) & \(0.384\pm0.0036\) & \(0.00209\pm0.0012\) & \(1.24\pm0.0019\) & ---\\
\end{longtable}

\begin{longtable}{llrrr}
\caption{Separate constraint components at horizon two, with three-seed descriptive spread. A dash means unavailable. SW2D height positivity is zero in all 108 learned-model seed/horizon records and is not repeated as rows of zeros here.}\label{tab:external_constraints}\\
\toprule
Task/model & Budget/filter & Density positivity & Pressure positivity & Magnetic divergence\\
\midrule
\endfirsthead
\toprule
Task/model & Budget/filter & Density positivity & Pressure positivity & Magnetic divergence\\
\midrule
\endhead
\bottomrule
\endfoot
CNS3D / FNO & Standard & \(0\pm0\) & \(0.00153\pm0.0011\) & ---\\
CNS3D / FNO & Short & \(0\pm0\) & \(0.0156\pm0.014\) & ---\\
CNS3D / FNO & Smoothed & \(0\pm0\) & \(0\pm0\) & ---\\
CNS3D / U-Net & Standard & \(0\pm0\) & \(9.72e-05\pm1.6e-05\) & ---\\
CNS3D / U-Net & Short & \(0\pm0\) & \(0.00186\pm0.0018\) & ---\\
CNS3D / U-Net & Smoothed & \(0\pm0\) & \(0\pm0\) & ---\\
CNS3D / ResNet & Standard & \(0\pm0\) & \(0.00839\pm0.0041\) & ---\\
CNS3D / ResNet & Short & \(0\pm0\) & \(0.00324\pm0.0011\) & ---\\
CNS3D / ResNet & Smoothed & \(0\pm0\) & \(2.34e-05\pm2.3e-05\) & ---\\
MHD3D / FNO & Standard & \(0\pm0\) & --- & \(0.263\pm0.0051\)\\
MHD3D / FNO & Short & \(0\pm0\) & --- & \(0.249\pm0.002\)\\
MHD3D / FNO & Smoothed & \(0\pm0\) & --- & \(0.0831\pm0.0044\)\\
MHD3D / U-Net & Standard & \(0\pm0\) & --- & \(0.255\pm0.0024\)\\
MHD3D / U-Net & Short & \(0\pm0\) & --- & \(0.237\pm0.003\)\\
MHD3D / U-Net & Smoothed & \(0\pm0\) & --- & \(0.0851\pm0.0019\)\\
MHD3D / ResNet & Standard & \(0\pm0\) & --- & \(0.254\pm0.0017\)\\
MHD3D / ResNet & Short & \(0\pm0\) & --- & \(0.288\pm0.01\)\\
MHD3D / ResNet & Smoothed & \(0\pm0\) & --- & \(0.0809\pm0.0013\)\\
\end{longtable}

\clearpage
\noindent\begin{minipage}{\linewidth}
\makeatletter\def\@captype{table}\makeatother
\centering
\caption{Half-step chains for Table~\ref{tab:rollout}'s checkpoints, retaining five seeds and crossed-bootstrap 95\% intervals. Errors/partition gaps are percentages; excess \(4T-T\) error is in percentage points across different reference times. The \(4T\) partition gap is normalized by the full-step prediction norm.}
\label{tab:rollout_half}
\resizebox{\linewidth}{!}{
\begin{tabular}{llrrrr}
\toprule
Equation & Training/selection & Half-step \(2T\) & Half-step \(4T\) & \(4T-T\) gap (pp) & Partition gap\\
\midrule
Heat & Supervised/best & \(0.064\,[0.055,0.072]\) & \(0.100\,[0.078,0.121]\) & \(0.006\,[-0.019,0.033]\) & \(0.125\,[0.093,0.175]\)\\
Heat & PINO-style/best & \(0.059\,[0.042,0.085]\) & \(0.092\,[0.055,0.151]\) & \(0.037\,[0.028,0.046]\) & \(0.097\,[0.074,0.119]\)\\
Heat & PINO-style/match & \(0.073\,[0.066,0.081]\) & \(0.096\,[0.088,0.103]\) & \(0.073\,[0.056,0.092]\) & \(0.118\,[0.092,0.144]\)\\
Heat & Composition/best & \(0.090\,[0.075,0.106]\) & \(0.122\,[0.089,0.157]\) & \(0.050\,[0.026,0.080]\) & \(0.068\,[0.055,0.078]\)\\
Heat & Composition/match & \(0.091\,[0.077,0.106]\) & \(0.124\,[0.091,0.158]\) & \(0.051\,[0.028,0.081]\) & \(0.067\,[0.053,0.078]\)\\
Advection & Supervised/best & \(0.333\,[0.284,0.385]\) & \(0.664\,[0.567,0.767]\) & \(0.663\,[0.540,0.796]\) & \(0.898\,[0.708,1.094]\)\\
Advection & PINO-style/best & \(0.299\,[0.235,0.379]\) & \(0.593\,[0.466,0.752]\) & \(0.373\,[0.300,0.463]\) & \(0.518\,[0.415,0.630]\)\\
Advection & PINO-style/match & \(0.477\,[0.384,0.579]\) & \(0.950\,[0.764,1.155]\) & \(0.584\,[0.487,0.691]\) & \(0.828\,[0.664,1.023]\)\\
Advection & Composition/best & \(0.425\,[0.369,0.495]\) & \(0.847\,[0.735,0.992]\) & \(0.709\,[0.632,0.807]\) & \(0.560\,[0.494,0.635]\)\\
Advection & Composition/match & \(0.454\,[0.380,0.541]\) & \(0.904\,[0.755,1.077]\) & \(0.764\,[0.660,0.885]\) & \(0.603\,[0.513,0.697]\)\\
Burgers & Supervised/best & \(0.089\,[0.072,0.106]\) & \(0.153\,[0.114,0.193]\) & \(0.039\,[0.012,0.063]\) & \(0.133\,[0.103,0.163]\)\\
Burgers & PINO-style/best & \(0.135\,[0.102,0.171]\) & \(0.272\,[0.145,0.398]\) & \(0.154\,[0.084,0.240]\) & \(0.268\,[0.194,0.361]\)\\
Burgers & PINO-style/match & \(0.145\,[0.113,0.178]\) & \(0.289\,[0.160,0.412]\) & \(0.161\,[0.079,0.267]\) & \(0.285\,[0.199,0.387]\)\\
Burgers & Composition/best & \(0.151\,[0.109,0.208]\) & \(0.290\,[0.168,0.442]\) & \(0.107\,[-0.016,0.257]\) & \(0.094\,[0.065,0.127]\)\\
Burgers & Composition/match & \(0.151\,[0.109,0.208]\) & \(0.290\,[0.168,0.442]\) & \(0.107\,[-0.016,0.257]\) & \(0.094\,[0.065,0.127]\)\\
\bottomrule
\end{tabular}

}
\end{minipage}

\clearpage
\noindent\begin{minipage}{\linewidth}
\makeatletter\def\@captype{table}\makeatother
\centering
\caption{Full-domain/octant mass balance at horizon two, including references. Entries give means and sample standard deviations of three seed means; overlapping windows preclude independent-trajectory intervals. The octant retains flux; periodic full-domain flux cancels.}
\label{tab:external_subvolumes}

\begin{tabular}{llrr}
\toprule
Task/model & Variant & Full-domain balance & Octant balance\\
\midrule
CNS3D/fno & Smoothed & \(0.01051\pm0.0048\) & \(0.01067\pm0.0042\)\\
CNS3D/fno & Short & \(0.01965\pm0.0035\) & \(0.0205\pm0.0054\)\\
CNS3D/fno & Standard & \(0.01051\pm0.0048\) & \(0.0107\pm0.004\)\\
CNS3D/reference & Reference & \(0.0001038\pm8.9e-06\) & \(0.003097\pm0.00015\)\\
CNS3D/resnet & Smoothed & \(0.004092\pm0.0026\) & \(0.005717\pm0.0011\)\\
CNS3D/resnet & Short & \(0.003564\pm0.002\) & \(0.003718\pm0.00089\)\\
CNS3D/resnet & Standard & \(0.004092\pm0.0026\) & \(0.005877\pm0.00078\)\\
CNS3D/unet & Smoothed & \(0.01384\pm0.0083\) & \(0.01362\pm0.0087\)\\
CNS3D/unet & Short & \(0.03412\pm0.02\) & \(0.03491\pm0.016\)\\
CNS3D/unet & Standard & \(0.01384\pm0.0083\) & \(0.01383\pm0.0083\)\\
MHD3D/fno & Smoothed & \(0.001603\pm0.00088\) & \(0.005039\pm0.001\)\\
MHD3D/fno & Short & \(0.003848\pm0.0027\) & \(0.006597\pm0.0026\)\\
MHD3D/fno & Standard & \(0.001603\pm0.00088\) & \(0.004805\pm0.00039\)\\
MHD3D/reference & Reference & \(0.0004289\pm0\) & \(0.004088\pm0\)\\
MHD3D/resnet & Smoothed & \(0.002094\pm0.0012\) & \(0.004069\pm0.0003\)\\
MHD3D/resnet & Short & \(0.002769\pm0.0016\) & \(0.004886\pm0.0019\)\\
MHD3D/resnet & Standard & \(0.002094\pm0.0012\) & \(0.005691\pm0.00039\)\\
MHD3D/unet & Smoothed & \(0.002446\pm0.0012\) & \(0.005311\pm0.00043\)\\
MHD3D/unet & Short & \(0.003116\pm0.00087\) & \(0.004811\pm0.0011\)\\
MHD3D/unet & Standard & \(0.002446\pm0.0012\) & \(0.003945\pm0.00066\)\\
\bottomrule
\end{tabular}

\end{minipage}
\end{landscape}

\section*{Declaration of generative AI and AI-assisted technologies in the manuscript preparation process}
During the revision stage, the author used OpenAI's Codex with GPT 6.1 Sol, GPT 6 Sol, and GPT 6 Astra to assist with manuscript editing and updates to the codebase, tests, experiment and analysis scripts, and reproducibility workflows. Anthropic's Claude Opus 5.5 provided feedback on the revision. The author directed and supervised this assistance and is responsible for reviewing and editing the assisted material and for the final content of this article.

\bibliographystyle{elsarticle-num}
\bibliography{references}

\end{document}